\documentclass[aps,nofootinbib,prd,preprintnumbers]{revtex4}
\usepackage{amstext}
\usepackage{amssymb}
\usepackage{amsmath}
\usepackage{graphicx}
\usepackage{rotating}
\usepackage{gensymb}
\usepackage[format=plain,singlelinecheck = false, format= hang, justification=raggedright]{caption}

\usepackage{multirow} 
\usepackage[T1]{fontenc} 
\usepackage{slashed}
\usepackage{caption}
\usepackage{float}
\usepackage[normalem]{ulem}

\usepackage[usenames,dvipsnames]{color}
\definecolor{mightnightblue}{RGB}{25,25,112}
\usepackage{hyperref}
 \hypersetup{
     colorlinks=true,
     linkcolor= Black,
     citecolor=mightnightblue,
     urlcolor=mightnightblue
     } 
\usepackage[utf8]{inputenc}

\def\gsim{\raise0.3ex\hbox{$\;>$\kern-0.75em\raise-1.1ex\hbox{$\sim\;$}}}
\def\lsim{\raise0.3ex\hbox{$\;<$\kern-0.75em\raise-1.1ex\hbox{$\sim\;$}}}
\newcommand {\ignore}[1]{}

\newcommand{\ie} {{\it i.e. }}
\newcommand{\eVq}  {\text{eV}^2}

\newcommand{\AddrAHEP}{%
  AHEP Group, Institut de F\'{i}sica Corpuscular --
  CSIC./Universitat de Val\`{e}ncia, Parc Cientific de Paterna.\\
 C/ Catedratico Jos\'e Beltr\'an, 2 E-46980 Paterna (Val\`{e}ncia) - SPAIN}

\begin{document}

\begin{flushright}
FTUV-18-04-13, IFIC/18-17 \\
\end{flushright}

\title{New physics vs new paradigms: distinguishing CPT violation from NSI}

\author{G. Barenboim$^1$}\email{Gabriela.Barenboim@uv.es}
\author{C. A. Ternes$^2$}\email{chternes@ific.uv.es}
\author{M. T{\'o}rtola~$^2$}\email{mariam@ific.uv.es}  
\affiliation{$^1$~Departament de F{\'i}sica Te{\'o}rica and IFIC, Universitat de Val{\`e}ncia-CSIC, E-46100, Burjassot, Spain}
\affiliation{$^2$~\AddrAHEP}


\vskip 2cm

\begin{abstract}
Our way of describing Nature is based on local relativistic quantum field theories, and then
CPT symmetry, a natural consequence of Lorentz invariance, locality and hermiticity of the Hamiltonian, is one of the few if not the only prediction that all of them share.
Therefore, testing CPT invariance does not test a particular model but the whole
paradigm. Current and future long baseline experiments will assess the status of CPT in the neutrino sector at an unprecedented
level and thus its distinction from similar experimental signatures arising from non-standard interactions is imperative. Whether the whole paradigm is at stake or just the 
standard model of neutrinos crucially depends on that.

\end{abstract}
\maketitle

\section{Introduction}
\label{sec:intro}
The status of the CPT symmetry as one of the cornerstones of particle physics has its origin on the fact that the theorem that protects it, the CPT Theorem~\cite{Streater:1989vi}, is one of the few solid predictions of any  local relativistic quantum field theory. 
This theorem states that particle and antiparticle possess the same mass and the same lifetime if they are unstable. The three requirements for the theorem to be proven are
\begin{itemize}
\item Locality
\item Lorentz invariance
\item Hermiticity of the Hamiltonian
\end{itemize}
which are key ingredients of our theories for reasons other than the status of the CPT
symmetry. Precisely because of that, testing CPT is testing our model building strategy and
our description of Nature in terms of local relativistic quantum fields.

If CPT is not conserved, one of the assumptions above has to be violated. 
Quantum gravity, for example, is expected to be non-local. However its effects are suppressed
by the Planck mass! But this is exactly the right ball-park for neutrinos to show it.
In fact, neutrinos are not only an ideal system to accommodate CPT violation~\cite{Barenboim:2002tz}
but also the most accurate tool to test it as well. In Ref.~\cite{Barenboim:2017ewj}
we have shown that the most stringent limits on CPT violation arise not from the kaon system but
from neutrino oscillation experiments. And contrary to the kaon case, these limits will improve 
in the next years significantly.
Given the impact such a result would have, it is crucial to distinguish a true, genuine CPT violation
from just a new, unknown interaction, no matter how sophisticated. After all, "outstanding claims require 
outstanding evidence".

In this work we will confront the presence of intrinsic CPT violation with new neutrino interactions with matter usually known as neutrino  non-standard interactions (NSI)~\footnote{Note that CPT violation  might also be induced through different new physics scenarios, such as decoherence~\cite{Gomes:2018inp,Capolupo:2018hrp,Carrasco:2018sca} or Lorentz violation~\cite{Kostelecky:2003cr,Diaz:2015dxa,Barenboim:2018ctx}. }.
NSIs are an agnostic  way to parametrize all the CPT preserving possibilities economically and therefore provide the ideal tool to  discriminate between a legitimate CPT violation, a phenomenon that challenges our description of Nature in terms of local relativistic quantum fields and a 
more or less complicated interaction that can be accommodated in the current paradigm. 
Here we will explore how such a distinction can be made if a different mass and/or mixing
pattern is extracted from the experiments when analyzing neutrino and antineutrino data sets separately, as it happens in the case of T2K, where different best fit points are obtained for  neutrino and antineutrino oscillation parameters~\cite{Abe:2017bay}.

Our paper is structured as follows: in Sections~\ref{sec:theo} and~\ref{sec:res-theo} we develop an analytic approach using the 2-neutrino approximation for illustration purposes, showing how CPT violation can mimic NSI. Next, in Section~\ref{sec:res-num} we present the details of the full simulation of DUNE using the the complete 3-neutrino picture. Our results are also presented and discussed in this section. Finally, we conclude in Section~\ref{sec:summary} with a summary.

\section{Theoretical background}
\label{sec:theo}

Neutrino non-standard interactions with matter are generically parameterized in terms of a low-energy effective four-fermion Lagrangian~\cite{Wolfenstein:1977ue,Valle:1987gv,Roulet:1991sm,Guzzo:1991hi}:
\begin{equation}
\mathcal{L}_{\rm NC-NSI} \, = \,
  - 2\sqrt{2} G_{F}  \, {\epsilon}^{f X}_{\alpha\beta}
      \left(\bar{\nu}_\alpha \gamma^{\mu} P_L \nu_\beta \right)
      \left(\bar{f}\gamma_{\mu} P_X f \right) \, ,
  \label{eq:NSI-Lagrangian}
\end{equation}
where  $P_X$ denotes the left and right chirality projection operators $P_{R,L} = (1\pm \gamma_5)/2$ and $G_F$ is the Fermi constant. The dimensionless coefficient $\epsilon_{\alpha\beta}^{f X}$ quantifies the strength of the NSI between  neutrinos of  flavour $\alpha$ and $\beta$ and the fundamental fermion $f\in\{ e, u,d \}$.

In the presence of such new non-standard interactions of neutrinos with matter, the effective two-neutrino Hamiltonian  governing neutrino propagation in the $\mu -\tau$ sector takes the form
\begin{equation}
H= \frac{1}{2E} \left[ U \left( {\begin{array}{cc}
   0 &  0 \\
    0 & \Delta m^2 \\
  \end{array} } \right) U^\dagger + A \left( {\begin{array}{cc}
   \epsilon^m_{\mu \mu} & \epsilon^{m*}_{\mu \tau} \\
  \epsilon^m_{\mu \tau}  & \epsilon^m_{\tau \tau}\\
  \end{array} } \right)
 \right],
 \label{eq:NSI_Hamiltonian}
\end{equation}
where
\begin{equation}
 U = \left( {\begin{array}{cc}
   \cos\theta & \sin\theta  \\
    -\sin\theta & \cos\theta \\
  \end{array} } \right)
\end{equation}
represents the leptonic mixing parametrized by the mixing angle in vacuum $\theta$. $\Delta m^2$ corresponds to the neutrino mass splitting in vacuum, $E$ is the neutrino energy
and $A= 2 \sqrt{2} G_F N_e E $ is the matter potential depending on the electron number density, $N_e$, along the neutrino trajectory. The parameters $\epsilon^m_{\mu \mu}$, $\epsilon^m_{\mu \tau}$ and $\epsilon^m_{\tau \tau}$ give the relative strength of the non-standard interactions compared to the Standard Model weak interactions. The superscript $m$ indicates that these parameters describe non-standard neutrino matter effects in a medium.
In this case, the relevant effective operator corresponds to the vector part of the interaction described in Eq.~(\ref{eq:NSI-Lagrangian}), namely
$\epsilon_{\alpha\beta}^{f V} = \epsilon_{\alpha \beta}^{f L} + \epsilon_{\alpha \beta}^{f R}$.
Considering a medium formed by first generation fermions, the effective NSI coupling in matter affecting neutrino propagation is given by
\begin{equation}
\epsilon^m_{\alpha \beta} \equiv \epsilon_{\alpha \beta}^{eV}+\frac{N_u}{N_e} \epsilon_{\alpha \beta}^{uV}+\frac{N_d}{N_e} \epsilon_{\alpha \beta}^{dV} \, ,
\label{eq:combination}
\end{equation}
with $N_f$ being the number density for the fermion $f\in\{ e, u,d \}$.
The flavor changing NSI parameters $\epsilon^m_{\alpha \beta}$ can in general be complex, while the flavor conserving coefficients $\epsilon^m_{\alpha \alpha}$ have to be real to preserve the hermiticity of the  Hamiltonian.
On the other hand, diagonal terms in the Hamiltonian proportional to the identity matrix do not affect the neutrino oscillation probability, and therefore one can subtract $\epsilon^m_{\mu \mu}$ from the diagonal in the Hamiltonian in Eq.~(\ref{eq:NSI_Hamiltonian}).   In consequence, oscillation experiments are only sensitive to the combination $\epsilon^m_{\tau \tau} -\epsilon^m_{\mu \mu}$. For simplicity, and given the existence of stronger constraints on the NSI couplings for muon neutrinos compared to tau neutrinos, in the following we will set $\epsilon^m_{\mu \mu}=0$.

\section{Analytical results at the probability level}
\label{sec:res-theo}

In the two--neutrino approximation, the neutrino survival probability for muon neutrinos in matter of constant density is given by
\begin{equation}
 P_{\mu\mu}=1-\sin^22\theta_m\sin^2\left(\frac{\Delta m_m^2 L}{4 E}\right)\,,
\end{equation}
 where $\theta_m$ and $\Delta m_m^2$ are the effective  mixing angle and effective mass squared difference in matter, respectively. 
 In the standard scenario, this probability is the same for neutrinos and antineutrinos.
However, this is not true if non-standard neutrino interactions are present, since they
 affect differently neutrinos and antineutrinos. This difference results basically in a shift in the effective neutrino and antineutrino oscillation parameters, so for neutrinos one would have
\begin{eqnarray}
 & \Delta m_\nu^2\cos 2\theta_\nu = \Delta m^2\cos 2\theta +\epsilon^m_{\tau\tau} A\,,
 \\
& \Delta m_\nu^2\sin 2\theta_\nu = \Delta m^2\sin 2\theta +2 \epsilon^m_{\mu\tau} A\,,
\end{eqnarray}
Similarly, for antineutrinos we write
\begin{eqnarray}
& \Delta m_{\overline{\nu}}^2\cos 2\theta_{\overline{\nu}} =\Delta m^2\cos 2\theta -\epsilon^m_{\tau\tau} A\,,\\
 & \Delta m_{\overline{\nu}}^2\sin 2\theta_{\overline{\nu}} =\Delta m^2\sin 2\theta -2  \epsilon^m_{\mu\tau} A\,,
\label{eq:diff1} 
\end{eqnarray}
 where we assume $\epsilon^m_{\mu\tau}$ to be real for simplicity. Here the subscript $\nu$ ($\overline{\nu}$) indicates  the effective mixing parameters in matter for neutrinos (antineutrinos). It is straightforward to see that  these equations  can be rearranged to express the unknown
parameters $\epsilon^m_{\mu\tau} A $, $\epsilon^m_{\tau\tau} A $, $\Delta m^2 $ and $\sin 2\theta$,
i.e. the "physical" parameters, in terms of the four observables $\Delta m_\nu^2$,
$\Delta m_{\overline{\nu}}^2 $, $\sin 2\theta_\nu$ and $\sin 2\theta_{\overline{\nu}}$:
\begin{eqnarray}
\label{eq:del1}
4 (\Delta m^2)^2  & = & (\Delta m_\nu^2)^2 + (\Delta m_{\overline{\nu}}^2)^2 + 2 
\Delta m_\nu^2 \Delta m_{\overline{\nu}}^2 \cos(2\theta_\nu-2\theta_{\overline{\nu}})\, , \\
\nonumber\\[-2mm]
\label{eq:del2}
\sin^2 2\theta & = & \frac{\left( \Delta m_\nu^2\sin 2\theta_\nu +  \Delta m_{\overline{\nu}}^2\sin 2\theta_{\overline{\nu}}\right)^2}{(\Delta m_\nu^2)^2 + (\Delta m_{\overline{\nu}}^2)^2 + 2 
\Delta m_\nu^2 \Delta m_{\overline{\nu}}^2 \cos(2\theta_\nu-2\theta_{\overline{\nu}} )}\, ,\\
\nonumber\\[-2mm]
2A\epsilon^m_{\tau\tau} & = & \Delta m_\nu^2\cos 2\theta_\nu -\Delta m_{\overline{\nu}}^2\cos 2\theta_{\overline{\nu}}\, , \label{eq:eps_mumu}\\
 4A\epsilon^m_{\mu\tau}& = & \Delta m_\nu^2\sin 2\theta_\nu -\Delta m_{\overline{\nu}}^2\sin 2\theta_{\overline{\nu}}\, .
 \label{eq:eps_mutau}
\end{eqnarray}

From these formulae it is obvious that a measurement of different neutrino and antineutrino mass splittings and/or mixing angles can in principle be explained by neutrino NSI with matter without resorting to CPT violation. 
Indeed, this idea was used to interpret different measurements in the neutrino and the antineutrino channel in the MINOS experiment~\cite{Kopp:2010qt}, although this result was not confirmed by more precise data from MINOS.
However, since there are already significant bounds
on the values of $\epsilon^m_{\tau\tau}$ and $\epsilon^m_{\mu\tau}$, it is also clear that there will be regions in the NSI parameter space which are excluded and then an experimental preference for such values would rather indicate a signal of CPT violation.

Going back to Eq.~(\ref{eq:eps_mumu}), if we assume  for simplicity the same mixing angles for neutrinos and antineutrinos~\footnote{Note that this is an approximation and, indeed, the effective mixing angles for neutrinos and antineutrinos are not equal, as shown in Eqs.~(\ref{eq:del1})-(\ref{eq:eps_mutau}). This simplification has been introduced to pedagogically illustrate the analogous role of the mass splittings and the NSI couplings in the CPT--violating and NSI scenario, respectively. An equivalent discussion can be done in terms of the mixing angles, assuming equal mass splittings for neutrinos and antineutrinos.}, $\theta_\nu=\theta_{\overline{\nu}}$, this equation can be rewritten to

\begin{equation}
\Delta(\Delta m^2) = \frac{2A\epsilon^m_{\tau\tau}}{\cos 2\theta}
 \label{eq:eps_mumu2b}
\end{equation}
with $\Delta(\Delta m^2)=\Delta m_\nu^2 - \Delta m_{\overline{\nu}}^2$ . Therefore, in the case of equal mixing angles, we obtain a very simple equation for $\Delta(\Delta m^2)$, which is linear in $\epsilon^m_{\tau\tau}$. In consequence, it is straightforward to interpret a different measurement of the neutrino and antineutrino mass splittings as caused by the presence of neutrino NSI with matter. Note that we have already calculated bounds on this difference in Ref.~\cite{Barenboim:2017ewj} from global oscillation data. However, no NSI couplings were considered there. Nevertheless, as mentioned before, there are experimental bounds available on the NSI couplings that should be taken into account. In Ref.~\cite{Farzan:2017xzy} one finds all the current bounds on the NSI parameters. Considering the definition of the effective NSI couplings in Eq.~(\ref{eq:combination}), the experimental bound on the diagonal NSI coupling is

\begin{equation}
 |\epsilon^m_{\tau\tau}|<0.11,\text{ at 90\% C.L.},
 \label{eq:bound_tt}
\end{equation}
  obtained in Ref.~\cite{GonzalezGarcia:2011my} from the analysis of atmospheric and MINOS long-baseline data. Assuming a gaussian distribution, we can extrapolate the previous bound to other confidence levels 

\begin{equation}
 |\epsilon^m_{\tau\tau}|< n(0.067),\text{ at $n\sigma$ C.L.}
\end{equation}
Therefore, in the presence of NSI, and according to Eq.~(\ref{eq:eps_mumu2b}), one can expect a difference in the effective values of the mass splitting measured in the neutrino and antineutrino channel.
The size of this deviation is  restricted by the existing bounds on the NSI couplings. For instance, at $n\sigma$ C.L., the maximum deviation will be given by

\begin{equation}
 \Delta(\Delta m^2) \leq \frac{2A}{\cos 2\theta}\times n(0.067) \, ,
 \label{eq:eps2}
\end{equation}
where we have taken

\begin{equation}
 A=2.27\times 10^{-4}\,\text{eV}^2\left(\frac{E}{\text{GeV}}\right)\,.
\end{equation}
That corresponds to the density of the Earth crust, $\rho$ = 3 $g/cm^3$.
This result also applies to the flavor violating NSI coupling $\epsilon^m_{\mu\tau}$. In this case, the current bound is given by

\begin{equation}
 |\epsilon^m_{\mu\tau}| <   0.018   \text{ at 90\% C.L.},
 \label{eq:bound_mt}
\end{equation}
originally given in Refs.~\cite{Esmaili:2013fva,Salvado:2016uqu} 
which translates to 

\begin{equation}
 |\epsilon^m_{\mu\tau}|< n(0.011),\text{ at $n\sigma$ C.L.}
\end{equation}
assuming the bound to be  gaussian. Therefore,  assuming equal mixing angles in both sectors, the deviation NSI can induce on the neutrino mass splitting over the antineutrino one can be expressed as

\begin{equation}
\Delta(\Delta m^2) = \frac{4A\epsilon^m_{\mu\tau}}{\sin 2\theta} \leq \frac{4A}{\sin 2\theta }\times n(0.011). 
 \label{eq:eps_mutau2b}
\end{equation}

\begin{figure}[t!]
 \centering
        \includegraphics[width=0.9\textwidth]{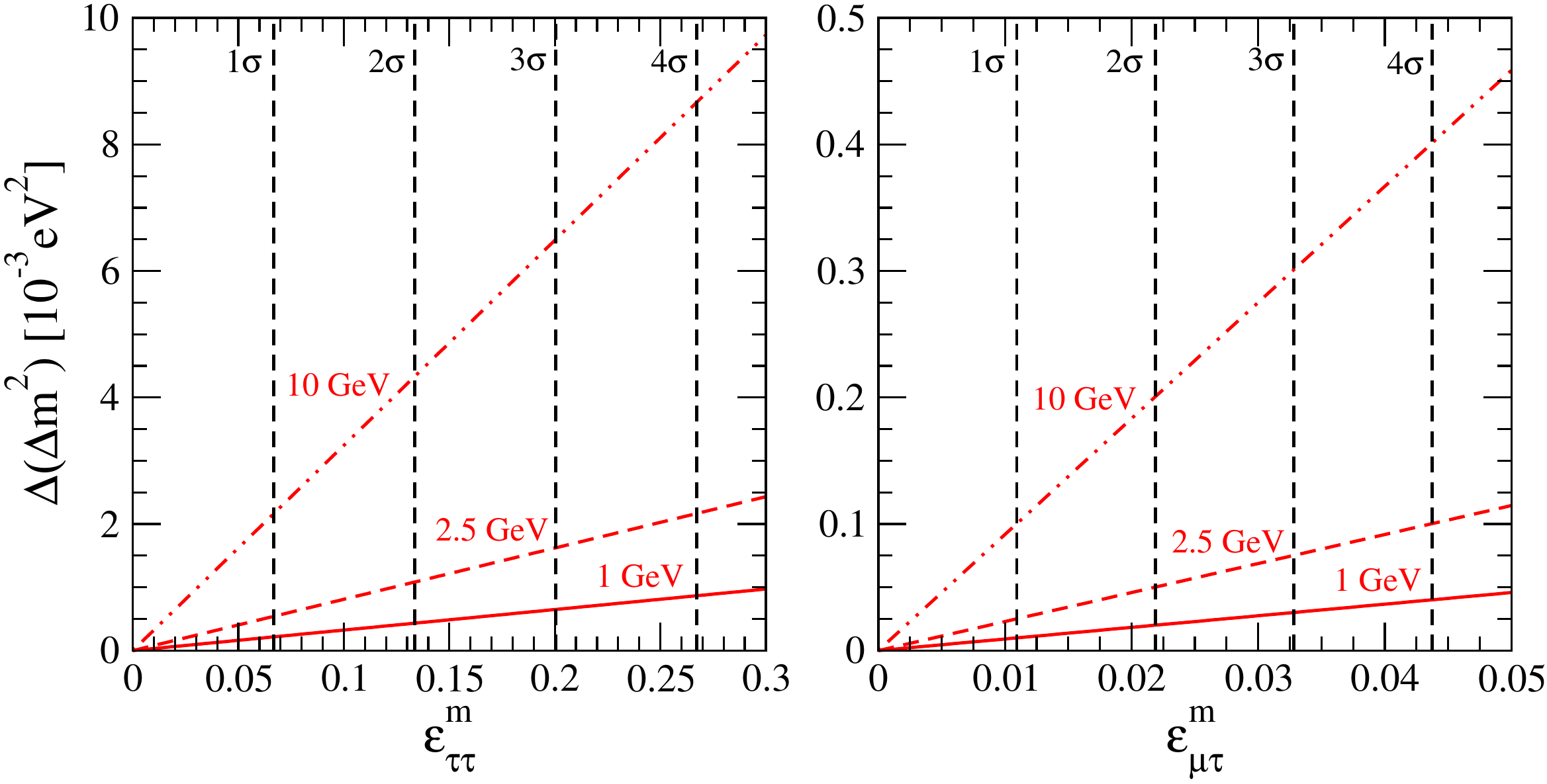}
       \caption{Left: $\Delta(\Delta m^2)$ as a function of $\epsilon_{\tau\tau}^m$, showing the allowed deviations of $\Delta(\Delta m^2)$ that 
       could be explained with NSI for $\theta=41\degree$ and $E=1.0$~GeV (solid), $E=2.5$~GeV (dashed) and $E=10$~GeV (dot-dashed line).  
        Right: The same for $\Delta(\Delta m^2)$ as a function of $\epsilon_{\mu\tau}^m$.}
	\label{fig:eps_ddm}
\end{figure}

The allowed deviations  between the neutrino and antineutrino mass splitting as a function of the NSI coupling $\epsilon^m_{\tau\tau}$ ($\epsilon^m_{\mu\tau}$) are shown in the left (right) panel of Fig.~\ref{fig:eps_ddm}.  There we have chosen $\theta=41\degree$, inspired by the best fit value for the atmospheric angle of the global fit of neutrino oscillation parameters in Ref.~\cite{deSalas:2017kay}, and the energies $E=1.0,2.5,10$~GeV, being 2.5~GeV the peak energy for $\nu_e$ appearance at DUNE, see Ref.~\cite{Acciarri:2015uup}. From this figure one can read to which amount a measurement of $\Delta(\Delta m^2)$ different from zero could be induced by NSI instead of being a signal of intrinsic CPT violation. The vertical dashed black lines indicate the current (1-4)$\sigma$ bounds on $\epsilon^m_{\tau\tau}$ and $\epsilon^m_{\mu\tau}$. The difference in the slope of the two graphs can be understood from two facts: first, the bound on $\epsilon_{\mu\tau}^m$ is  stronger than the one for $\epsilon_{\tau\tau}^m$ and second, for the mixing angle of choice, we have $\cos 2\theta \approx 0.14$ and $\sin2\theta\approx0.99$, so the deviation potentially induced by $\epsilon_{\tau\tau}^m$ can be much larger.
Note that in this section we have assumed equal mixing angles, so that the situation could be even more confusing if one allows for different mixing angles in the neutrino and antineutrino sector. Notice also that these results have been derived at the probability level and, in principle, one can expect the picture to be somehow blurred    when the complete analysis  of a neutrino experiment is carried out taking into account the full simulated neutrino spectrum and the associated statistical and systematical errors.

\section{Results from the  simulation of DUNE}
\label{sec:res-num}

In Ref.~\cite{Barenboim:2017ewj} we have  shown that, if the results from the separate neutrino and antineutrino analysis of the T2K collaboration~\cite{Abe:2017bay} turn out to be true, DUNE could measure CPT violation at more than 3$\sigma$ confidence level. In this section we will analyze if indeed these results could be confused with NSI. 

The next generation experiment DUNE will be the so far biggest long baseline neutrino experiment. It will consist of two detectors exposed to a megawatt-scale muon neutrino beam produced at Fermilab. The near detector will be placed near the source of the beam, while a second, much larger, detector comprising four 10-kiloton liquid argon TPCs will be installed 1300 kilometers away of the neutrino source. One of the main scientific goals of DUNE is the precision measurement of the neutrino oscillation parameters. In our work, the simulation of DUNE is performed with the GLoBES package~\cite{Huber:2004ka,Huber:2007ji} with the most recent DUNE configuration file provided by the experimental collaboration~\cite{Alion:2016uaj}. The expected NSI signal in DUNE is simulated using the GLoBES extension \textit{snu.c}~\cite{Kopp:2006wp, Kopp:2007ne}. We assume a total run period of 3.5 years in the neutrino mode and another 3.5 years in the antineutrino mode. Assuming an 80~GeV beam with 1.07~MW beam power, this corresponds to an exposure of 300 kton--MW--years, which corresponds to $1.47\times 10^{21}$ protons on target (POT) per year, which amounts basically in one single year to the same statistics accumulated by T2K in all of its lifetime in runs 1--7c. Since here we focus basically on the parameters relevant for $P_{\mu\mu}$, we consider only the disappearance channel in our simulation. We include signals and backgrounds, where the latter include contamination of antineutrinos (neutrinos) in the neutrino (antineutrino) mode, and also misinterpretation of flavors. We have increased the systematic errors due to misidentification of neutrinos by antineutrinos and vice versa by a further 25\% over the original error given by the collaboration  to account for the fact that, actually, antineutrino backgrounds should not be oscillated with the neutrino oscillation parameters and vice versa. While it is possible to define a customized probability engine in GLoBES, we have checked that the effect of the backgrounds is not appreciable in the final results.

\begin{table}[t]\centering
   \begin{tabular}{lc}
    \hline
    parameter & value 
    \\
    \hline
    $\Delta m_{31}^2$ & 2.60$\times 10^{-3} \,\eVq$ \\
    $\Delta\overline{m}_{31}^2$  &2.62$\times 10^{-3} \,\eVq$ \\ 
    $\sin^2\theta_{23}$ & 0.51 \\
$\sin^2\overline{\theta}_{23}$ & 0.42\\[2mm]
\hline\\[-3mm]
    $\Delta m^2_{21}$, $\Delta\overline{m}^2_{21}$& $7.56\times 10^{-5} \,\eVq$\\  
    $\sin^2\theta_{12}$, $\sin^2\overline{\theta}_{12}$ & 0.321\\ 
     $\sin^2\theta_{13}$, $\sin^2\overline{\theta}_{13}$& 0.02155\\
   $\delta$, $\overline{\delta}$  & 1.50$\pi$\\
       \hline
     \end{tabular}
     \caption{Oscillation parameters considered in the simulation and analysis of DUNE expected data.
  In terms of the two neutrino parameterization of Sec.~\ref{sec:theo}, the notation used in this table is equivalent to $\Delta m_{31}^2 = \Delta m^2_\nu$, $\Delta\overline{m}^2_{31} = \Delta m^2_{\overline{\nu}}$ and $\theta_{23} = \theta_\nu$, $\overline{\theta}_{23} = \theta_{\overline{\nu}}$.}
     \label{tab:par1} 
\end{table}

Our strategy is as follows: we perform two simulations of DUNE running 3.5 years only in neutrino mode and 3.5 years only in antineutrino mode. To generate the future data, we consider the parameters presented in Tab.~\ref{tab:par1}, assuming different atmospheric mixing angle and mass splitting for neutrinos and antineutrinos (using the best fit points from T2K, see Tab.~\ref{tab:par1}), but no NSI. Then, in the statistical analysis, we scan over the standard oscillation parameters $\delta$, $\theta_{13}$, $\theta_{23}$, $\Delta m_{31}^2$ and their antineutrino counterparts\footnote{We put a prior on $\overline{\theta}_{13}$, due to the precise measurements by the short baseline reactor experiments~\cite{An:2016ses,RENO:2015ksa,Abe:2014bwa}.}. Additionally, we scan over the two NSI parameters relevant for our channel of interest, $\epsilon^m_{\tau\tau}$ and $\epsilon^m_{\mu\tau}$, see Eq.~(\ref{eq:NSI_Hamiltonian}). The remaining parameters are set to zero, since they only contribute  to $P_{\mu\mu}$ and $\overline{P}_{\mu\mu}$ at subleading orders. The same argument applies to the phase of $\epsilon_{\mu\tau}^m$, which is therefore set to zero, too. Summarizing, we simulate DUNE fake data under the assumption of CPT violation and then we perform the reconstruction of the neutrino signal assuming CPT conservation and the presence of  NSI with matter along the neutrino propagation.

\begin{figure}[t!]
 \centering
        \includegraphics[width=0.9\textwidth]{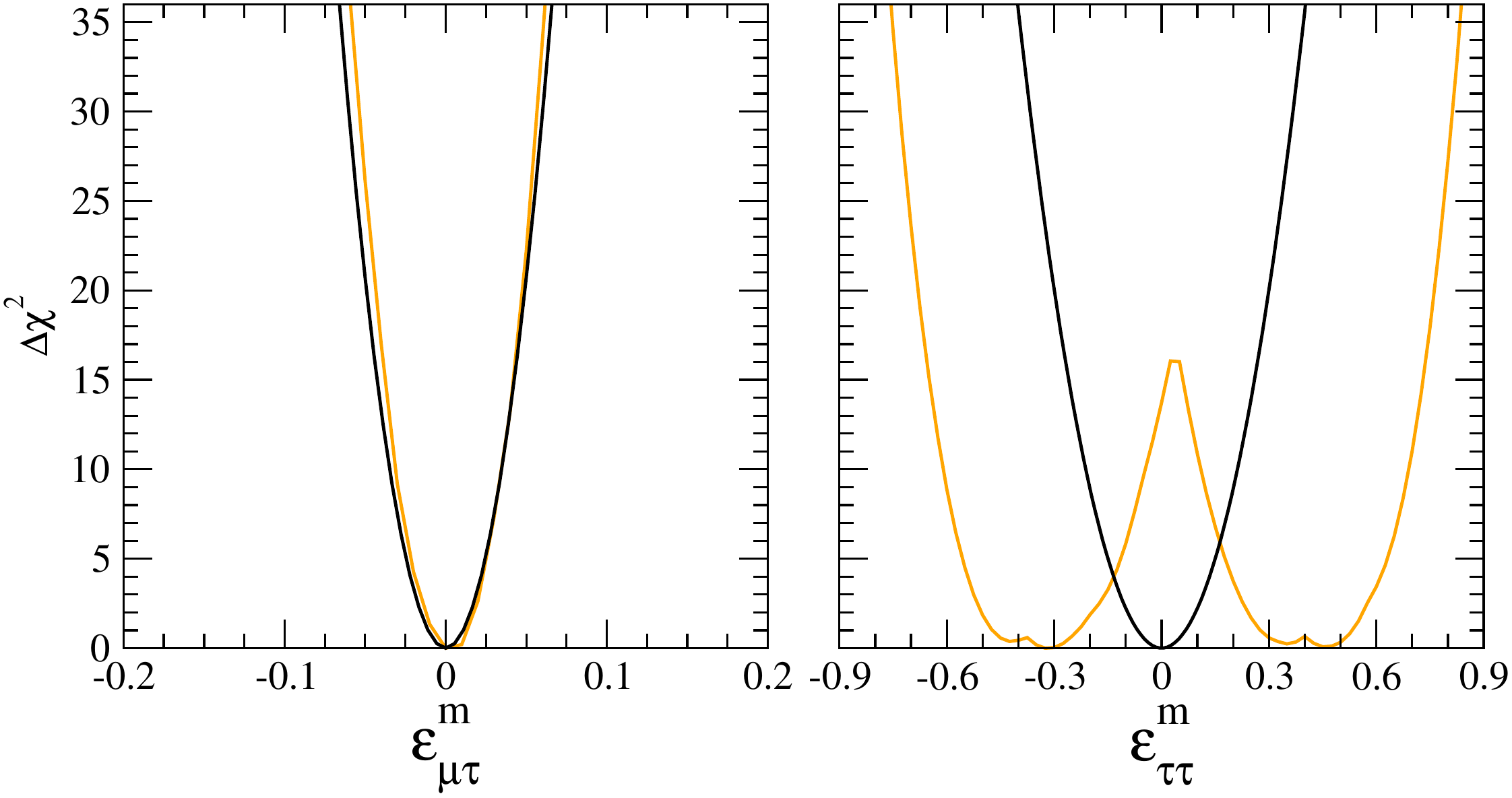}   
           \caption{$\Delta\chi^2$ profile for the NSI couplings  $\epsilon^m_{\mu\tau}$ (left) and $\epsilon^m_{\tau\tau}$ (right panel) 
           obtained in our analysis (orange lines). For comparison we show the current experimental bounds (see Eqs.~(\ref{eq:bound_tt}) and~(\ref{eq:bound_mt}) and the corresponding discussions) on both parameters assuming gaussian errors (black lines).}
	\label{fig:eps_mt-eps_mm-bound}
\end{figure}

Using this procedure, we obtain two $\chi^2$ grids, one for the neutrino mode and another one for the antineutrino mode. The total $\chi^2$  distribution is obtained  from the sum of the neutrino and antineutrino disappearance contributions\footnote{In our simulation we have assumed 71 energy bins between 0 and 20 GeV.}

\begin{equation}
 \chi^2_{\text{total}}=\chi^2(\nu)+\chi^2(\overline{\nu}) \, ,
\end{equation}
and it is marginalized over all the parameters except the one of interest. 
First of all, we focus our attention to the NSI parameter $\epsilon^m_{\mu\tau}$. The  $\Delta\chi^2$ profile obtained for this coupling is shown in the orange  line  in the left panel of Fig.~\ref{fig:eps_mt-eps_mm-bound}. 
Here we see that data are consistent with this NSI coupling equal to zero. 
However, the result is different for the flavour diagonal NSI coupling $\epsilon^m_{\tau\tau}$. The $\Delta\chi^2$ profile for this parameter is plotted in the orange line  in the right panel of Fig.~\ref{fig:eps_mt-eps_mm-bound}. Note that the black lines represent the current bounds on these quantities. From the figure we see our analysis  excludes the value $\epsilon^m_{\tau\tau}=0$ at approximately 4$\sigma$, while the best fit is obtained for $\epsilon^m_{\tau\tau}=-0.33$. 
Besides the absolute minimum, three other local minima are obtained with $\Delta\chi^2 < 0.4$. This multiplicity is due to 
 the octant degeneracy present in the analysis of the disappearance channel alone. As it is well known, disappearance data are mostly sensitive to $\sin^2 2\theta_{23}$ and, therefore, can not distinguish  the correct octant of $\theta_{23}$. In this case, the degeneracy in the determination of the atmospheric mixing angle appears twice, for the neutrino and the antineutrino mixing angle, and two different values of $\cos 2\theta_\nu$ and $\cos 2\theta_{\overline{\nu}}$ are preferred at the same confidence level. As a result, and as it can be easily understood from  Eq.~(\ref{eq:eps_mumu}),  four different values of $\epsilon_{\tau\tau}^m$  are allowed at roughly the same confidence level.
This degeneracy could be partially broken by including  the appearance channel in our analysis. However, doing this we would also have to include more NSI parameters in the analysis, since other NSI parameters as $\epsilon^m_{e \mu}$  and $\epsilon^m_{e \tau}$ would be very relevant for the   appearance channel.

\begin{figure}[t!]
 \centering
        \includegraphics[width=0.9\textwidth]{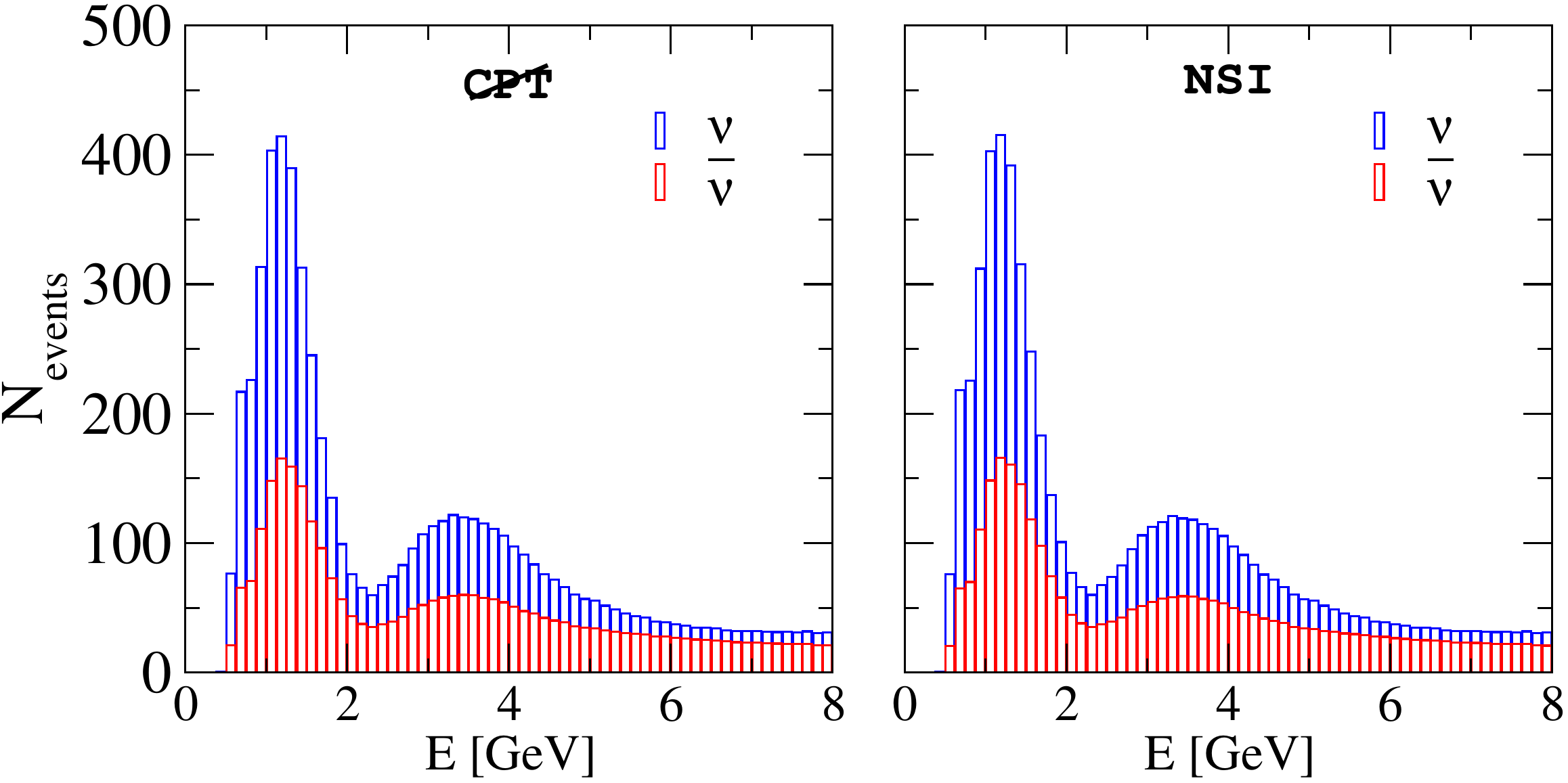}   
           \caption{Expected number of muon neutrino (blue) and muon antineutrino (red) events for CPT violation (left) and for the CPT--conserving scenario with NSI (right) at the DUNE experiment.
           The parameters used to create the plot in the CPT--violating case are the ones measured by T2K, summarized in Tab.~\ref{tab:par1}. For the NSI case, we used the combined new best fit, namely $\epsilon_{\tau\tau}^m=-0.33$, $\sin^2\theta_{23}^{\text{ comb}}=0.56$, $\sin^2\theta_{13}^{\text{ comb}}=0.0216$, $\Delta m_{31}^{2\text{ comb}}=2.62\times 10^{-3} \,\eVq$ and $\delta^{\text{ comb}} = 1.4\pi$. Note that also higher energy bins were included in the analysis, but are not shown here since the spectrum goes to zero for higher energies.}
	\label{fig:Events}
\end{figure}

Our results can be explained from the naive formulae in Eqs.~(\ref{eq:eps_mumu}) and (\ref{eq:eps_mutau}). Indeed, if we substitute the T2K best fit values for the oscillation parameters given in Table \ref{tab:par1}, used to simulate future DUNE data, into these formulae,  for  the peak energy of DUNE, $E=2.5$~GeV, we obtain $\epsilon^m_{\mu\tau}=5.8\times10^{-3}$ and $\epsilon^m_{\tau\tau}=-0.42$. Therefore, one can conclude that what seems to be CPT violation could be also explained with neutrino NSI with matter. 
 This can also be illustrated through the energy distribution of the number of expected events in DUNE for both scenarios, as shown in Fig.~\ref{fig:Events}. 
The left panel corresponds to the event distribution for neutrinos (blue) and antineutrinos (red), assuming CPT violation, with the oscillation parameters for neutrinos and antineutrinos given  in Tab.~\ref{tab:par1}.
The right panel, on the contrary, shows the spectrum of events expected in the CPT--conserving case with NSI.
It should be noted that, for this last scenario, we perform a combined analysis of the neutrino and antineutrino channel assuming equal values for the oscillation parameters. As a result, we not only obtain a non-zero coefficient $\epsilon_{\tau\tau}^m$, but also different standard oscillation parameters from the ones used to simulate the neutrino signal in DUNE\footnote{We had already discussed this type of imposter solutions in Ref.~\cite{Barenboim:2017ewj}.}. Indeed, a new best fit point is found, located  at $\epsilon_{\tau\tau}^m=-0.33$, $\sin^2\theta_{23}^{\text{ comb}}=0.56$, $\sin^2\theta_{13}^{\text{ comb}}=0.0216$, $\Delta m_{31}^{2\text{ comb}}=2.62\times 10^{-3} \,\eVq$ and $\delta^{\text{ comb}} = 1.4\pi$. 
 As one can see, both panels are almost identical, showing the equivalence between the two scenarios under analysis. 
 As a final remark, we would like to add that the two analyzed possibilities  can not be distinguished by means of $\chi^2$ either. Since we create fake data under the assumption of CPT violation,  the best fit point for the independent analysis of neutrino and antineutrino data automatically gets the value $\chi^2 = 0$. However, when we perform a CPT-conserving analysis with NSI,  the new best fit point has a value of only $\Delta\chi^2 = 0.55$ and, therefore,  one can not claim the fit to data is significantly worse in that case.

Nevertheless, it should be noted that the best fit value we have obtained in this NSI case for the flavor diagonal coupling, $\epsilon^m_{\tau\tau}=-0.33$, is  highly excluded from current experimental data\footnote{Note that most bounds in the literature are calculated by taking one or two NSI parameters at the time. Obviously, if one allows for several of them to be non-zero, much weaker bounds can be obtained, see for example Ref.~\cite{Esteban:2018ppq}.}.
To visualize this, together with the results  obtained in our analysis, we have plotted in Fig.~\ref{fig:eps_mt-eps_mm-bound} the profile for both NSI parameters from current data assuming gaussian errors (see black lines in both panels). In the right graph one can see that our preferred value for $\epsilon^m_{\tau\tau}$  is actually excluded at close to 5$\sigma$. Therefore, if the results from T2K (that we take as input parameters for the simulation of the CPT-violating data sample in DUNE)  turn out to be true, they would rather hint towards CPT violation than to NSI.

\section{Summary}
\label{sec:summary}

The impact of a potential CPT violation is such that it is imperative to distinguish it from
any other unknown physics that can lead to similar experimental signatures. The need for such a distinction is more urgent in the neutrino sector where the future long baseline neutrino experiments will push the CPT invariance frontier to a new level and where neutrinos, due to its uncommon mass generation mechanism, can open unique windows to new physics and new mass scales.

On  the other side, NSIs are a simple way to parametrize any unknown physics which may be relevant to neutrino oscillations. In this work we have shown that, although NSIs can mimic fake CPT violation, its experimental signature can be distinguished from a genuine CPT violation due to the bounds
on the strength of the NSI arising from other experiments. 
Indeed, we have found that the different results for the neutrino and antineutrino parameters measured by the T2K Collaboration may be interpreted in terms of a CPT--conserving scenario in combination with neutrino NSI with matter. The match between the prediction of both scenarios is astonishing, as shown in Fig. ~\ref{fig:Events}. However, this equivalence has a caveat, since the size of the diagonal NSI coupling required, $\epsilon_{\tau\tau}^m \simeq -0.3$, is excluded by current neutrino oscillation data.
Therefore, the future cannot be more exciting. If the slight indications of CPT violation in the mixing angles offered by the separate analysis of the  neutrino and antineutrino data sets by  T2K  are confirmed by the DUNE experiment, genuine CPT violation \ie  the one that challenges
our understanding of Nature in terms of local relativistic quantum field theory will be the only answer.
If this were not the case, we explain how the discrimination has to be performed in case a difference is ever found.

\section{Acknowledgments}

GB acknowledges support from the MEC and FEDER (EC) Grants SEV-2014-0398, FIS2015-72245-EXP, and FPA-2017-84543P  and the Generalitat Valenciana under grant PROMETEOII/2017/033. GB acknowledges partial support from the European Union FP7 ITN INVISIBLES MSCA PITN-GA-2011-289442 and InvisiblesPlus (RISE) H2020-MSCA-RISE-2015-690575. 
CAT and MT are  supported by the Spanish grants FPA2017-85216-P and
SEV-2014-0398 (MINECO) and PROMETEO/2018/165 and GV2016-142 grants
from Generalitat Valenciana.
CAT is supported by the FPI fellowship BES-2015-073593 (MINECO).
MT acknowledges financial support from MINECO through the Ram\'{o}n y Cajal contract RYC-2013-12438 as well as from the L'Or\'eal-UNESCO \textit{For Women in Science} initiative.

\begingroup
\raggedright
\sloppy


\end{document}